\documentclass[twocolumn,showpacs,showkeys,preprintnumbers,amsmath,amssymb]{revtex4}  

\usepackage{graphicx}
\usepackage{dcolumn}
\usepackage{bm}
\usepackage{multirow}
\usepackage{color}

\begin{document}

\title{Dark-bright mixing of interband transitions\\ in symmetric semiconductor quantum dots }

\author{G.~Sallen$^1$}
\author{B.~Urbaszek$^1$}
\email[Corresponding author : ]{urbaszek@insa-toulouse.fr}
\author{M.~M.~Glazov$^2$}
\author{E.~L.~Ivchenko$^2$}
\author{T.~Kuroda$^3$}
\author{T.~Mano$^3$}
\author{S.~Kunz$^1$}
\author{M.~Abbarchi$^3$}
\author{K.~Sakoda$^3$}
\author{D.~Lagarde$^1$}
\author{A.~Balocchi$^1$}
\author{X.~Marie$^1$}
\author{T.~Amand$^1$}

\affiliation{%
$^1$Universit\'e de Toulouse, INSA-CNRS-UPS, LPCNO, 135 Av. Rangueil, 31077 Toulouse, France}

\affiliation{%
$^2$Ioffe Physical-Technical Institute RAS, 194021 St.-Petersburg, Russia}

\affiliation{%
$^3$National Institute for Material Science, Namiki 1-1, Tsukuba 305-0044, Japan}

\date{\today}

\begin{abstract}
In photoluminescence spectra of symmetric [111] grown GaAs/AlGaAs quantum dots
in longitudinal magnetic fields applied along the growth axis we observe in addition to the expected bright states also nominally dark transitions for both charged and neutral excitons. We uncover a strongly non-monotonous, sign changing field dependence of the bright neutral exciton splitting resulting from the interplay
between exchange and Zeeman effects. Our theory shows quantitatively that these surprising experimental results are due to magnetic-field-induced $\pm 3/2$ heavy-hole mixing, an inherent property of systems with C$_{3v}$ point-group symmetry.

\end{abstract}

\pacs{72.25.Fe,73.21.La,78.55.Cr,78.67.Hc}
                            \keywords{Quantum dots, optical selection rules}
\maketitle
The search of methods to generate and manipulate entangled
quantum states is one of the driving forces behind experimental physics on
the nano-scale. The initial proposal to use the exciton-biexciton
cascade in quantum dots to generate entangled photon pairs
\cite{Benson2000} relies on symmetric dots where the
neutral exciton X$^0$ 
states  are degenerate, i.e. have zero
fine structure splitting 
$\delta_1$ induced by anisotropic electron-hole Coulomb exchange. As in practice $\delta_1\ne 0$ in the majority of quantum
dot systems \cite{Gammon1996a,Bayer1999,Besombes2000}, very inventive research has been
developed trying to tune the fine structure splitting to zero with original techniques
\cite{Stevenson2006,Plumhof2011,Langbein2004,Akopian2006}.  An
alternative, recent approach is to use samples grown along the $z' \parallel [111]$
crystallographic axis, which is also the orientation of most
nano-wires \cite{Lu2006}.  This growth axis has the advantage of providing
microscopically identical interfaces for quantum well or dot
structures, resulting in $C_{3v}$ point symmetry. Hence,
small fine structure 
splittings in \textit{as grown} [111] quantum dot
structures have been recently
predicted \cite{Schliwa2009,Singh2009}  and observed
\cite{Mano2010,Karlsson2010,Stock2010} followed by a first report of photon
  entanglement \cite{Mohan2010}.

\begin{figure}
\includegraphics[width=0.44\textwidth]{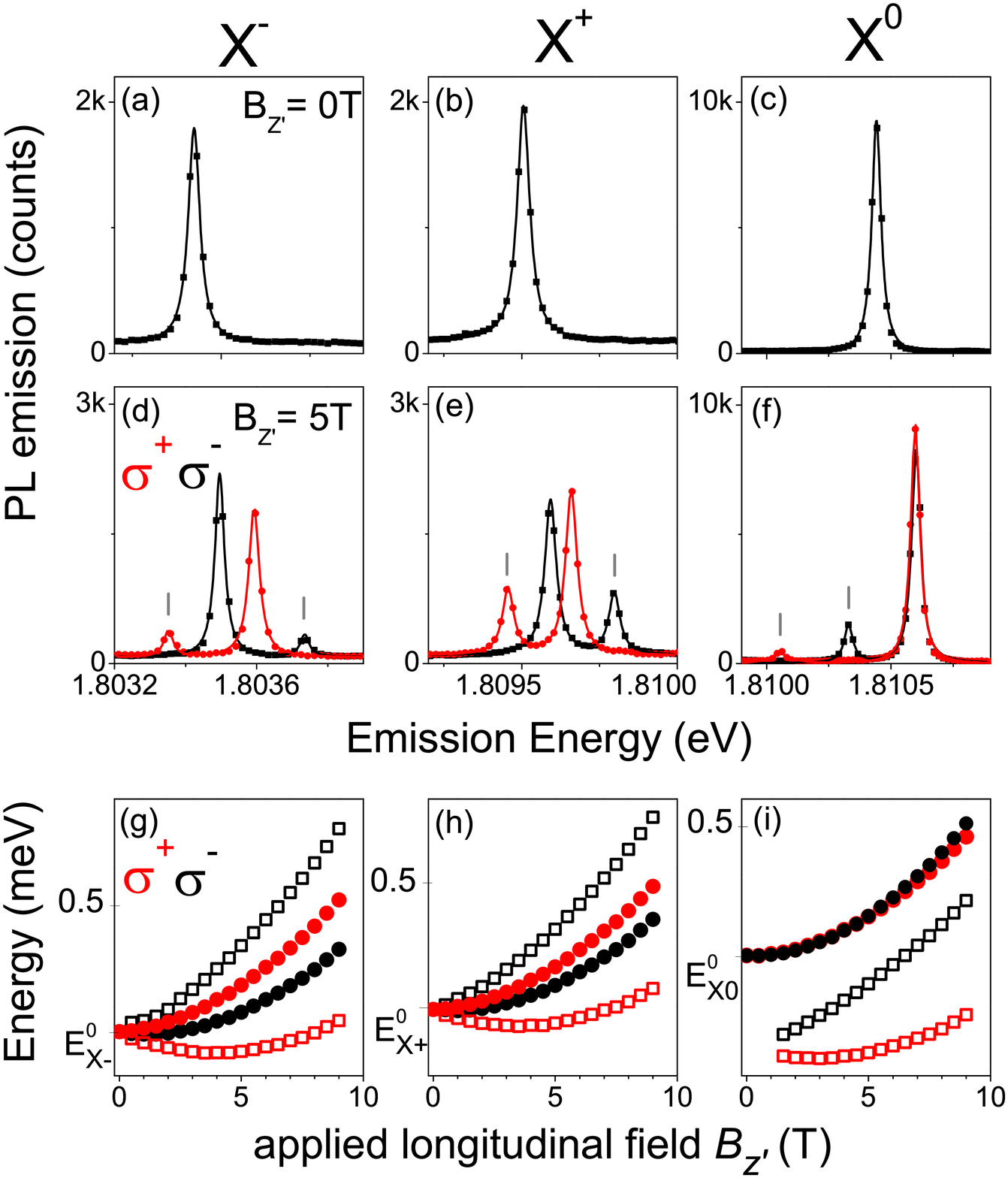}
\caption{\label{fig:fig1} (Color online): (a)-(c) single
  dot PL spectra at $B_{z'}=0$. (d)-(f) single dot PL spectra at
  $B_{z'}=5T$, for $\sigma^-$ (black line/solid squares) and
  $\sigma^+$ (red line/circles) (g)-(i) transition energies as a
  function of $B_{z'}$,  for $\sigma^-$: dark (black hollow squares)
  and bright (black circles) and $\sigma^+$: (red hollow squares) and
  bright (red circles). Data are shown for
    \textbf{QD I}. 
}
\end{figure}

In the commonly studied dot samples grown along $z\parallel
  [001]$ axis, the exact 
nature and symmetry of the X$^0$ and charged
exciton states is deduced from experiments in
longitudinal magnetic 
fields i.e. parallel to the growth axis. These studies made
  crucial contributions to the development of 
quantum dot photonics and spin 
physics \cite{Bayer1999,Besombes2000,Paillard2001} .  
Here we study the effect of a longitudinal magnetic field $\bm
  B\parallel z' \parallel [111]$  in strain free [111] grown GaAs quantum
  dots. In $B_{z'}\ne 0$
we observe \textit{four} emission lines, as two nominally
dark transitions emerge in 
addition to the usual bright Zeeman doublet for charged excitons and
X$^0$ of all quantum dots investigated. Our
measurements show that the heavy hole states with 
  spin projections $+3/2$ or $-3/2$ onto the growth axis $z'$ are coupled in a
longitudinal magnetic field.  
The resulting appearance of forbidden charged exciton and dark X$^0$ transitions due to hole mixing is an
inherent feature of [111] grown dots and is not related to a symmetry lowering, of the confinement potential or due to strain, as
in [001] grown dots \cite{Besombes2000,Bayer1999}.   
We are able to measure the dark-bright X$^0$ separation and observe a
strongly non-monotonous bright X$^0$ splitting that changes sign as a function of 
$B_{z'}$ due to the competition between isotropic electron-hole 
Coulomb exchange and the Zeeman interactions. 
 We explain all these intriguing findings by a
   microscopic model of the Zeeman interaction accounting for $C_{3v}$
   point symmetry of the studied quantum dots. 
The key
ingredient of our theory is to include the cubic terms for the hole hamiltonian \cite{Ivchenko1995,Marie1999} going beyond the
commonly used spherical approximation. 
We extract sign and magnitude of electron and hole $g$ factors for this
new system.

The sample was grown by droplet epitaxy using a conventional molecular
beam epitaxy system \cite{Mano2010,Abbarchi2010,Belhadj2008} on a GaAs(111)A substrate. 
The dots are grown on 100nm thick Al$_{0.3}$Ga$_{0.7}$As  barriers and are covered by 50nm of the same material.
There is no continuous wetting layer in the sample connecting the
dots (typical height $\simeq$3nm, radius $\simeq$15nm), see details in \cite{Mano2010}. Single dot photoluminescence (PL) at 4K is recorded with
a home build confocal microscope with a detection spot diameter of
$\simeq 1\mu m$. The detected PL polarization is analysed and the
signal is dispersed by a spectrometer and detected by a Si-CCD
camera. Optical excitation is achieved by pumping the AlGaAs barrier
with a HeNe laser at 1.96~eV that is linearly polarized to exclude the
effects of optical carrier orientation and dynamic nuclear
polarization \cite{belhadj09}. 

\begin{figure}
\includegraphics[width=0.45\textwidth]{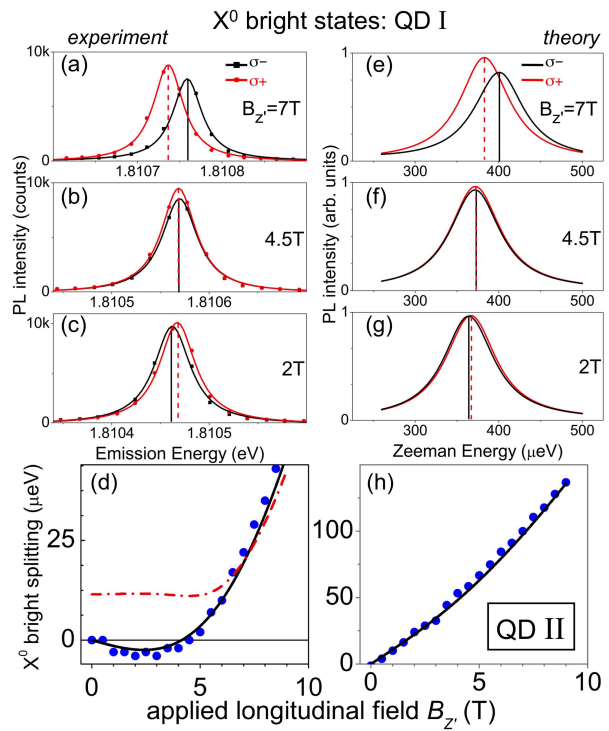}
\caption{\label{fig:fig2} (Color online):  (a)-(c)  PL
    spectra in
  $\sigma^{+/-}$ polarizations of bright X$^0$ for different
  $B_{z'}$. (d) bright X$^0$ Zeeman splitting $E(\sigma^-)-E(\sigma^+)$ vs $B_{z'}$: experiment
  (circles), theory (black line), theoretical value of total splitting
   including $\delta_1=11 \mu eV$ (dotted red
  line); (e)-(g) calculated spectra. Panels
    (a)-(g) correspond to \textbf{QD I}. (h) as (d) but for \textbf{QD II}. 
} 
\end{figure}

Figure 1a-c shows the different emission lines originating from a
typical quantum dot QD I at zero magnetic field. The X$^0$, the negatively charged exciton  X$^-$ (2
electrons, 1 hole) and the positively charged exciton X$^+$ (1
electron, 2 holes) are identified using fine structure analysis and
optical orientation experiments \cite{belhadj09}. The high symmetry of the dots is
reflected in typical values for the splitting of the X$^0$ emission
due to anisotropic electron-hole exchange $\delta_1$ of a few $\mu$eV
\cite{Mano2010}, extracted from angle dependent PL polarization
analysis in the linear basis.  

In figure 1d-f the $\sigma^+$ and $\sigma^-$ polarized emission from
the same exciton states are presented in the presence of a
longitudinal magnetic field $B_{z'}=5T$. 
We first discuss charged excitons X$^+$ and X$^-$, whose
  emission energies are shown in Fig.~1g,h. In contrast to the
widely studied [001] grown samples, where
a Zeeman doublet is observed, with
one $\sigma^+$ and one $\sigma^-$ polarized branch
\cite{Leger2007,Abbarchi2010,Belhadj2008,Bayer1999} here the emission
patterns 
are strikingly different: We observe in total 
four transitions, two of them are $\sigma^+$ polarized,
and two others are $\sigma^-$ polarized. 
For each polarization, the more intense emission line will be called
"bright", the less intense "dark" in the following. 
The emission of two doublets is observed for the X$^+$ and the X$^-$
exciton of all the dots studied as soon as $|B_{z'}|>0$ in this sample, see
Fig.~1g,h. 
The measured ratio of the emission intensity bright
/dark transitions remains constant as $|B_{z'}|$ changes (not shown).  

We also note the appearance of dark states for the X$^0$
emission in Fig. 1f,i. For typically
  $|B_{z'}|>2T$ we are able to 
detect that the bright X$^0$ emission is accompanied by less intense
lines at $\delta_0\simeq 350 \mu$eV  lower in energy. This energy  separation $\delta_0$ is
due to isotropic electron-hole exchange which splits bright and dark states.
Previously, dark X$^0$ states have been
observed generally for dots grown along the
[001] axis either in high \textit{transverse} magnetic fields (Voigt geometry) \cite{Puls1999} or exceptionally
in high \textit{longitudinal} magnetic fields for dots with lowered symmetry \cite{Leger2007,Bayer1999}. 
In the dots grown along [111] investigated here the 
dark X$^0$ is clearly visible for \textit{all} dots in this sample in
  the Faraday configuration, even for highly symmetric dots with vanishing $\delta_1$.   

Another surprising feature of the X$^0$ emission is shown in
Fig. 1i and analyzed in detail in Fig. 2a-c. At 2T, the $\sigma^+$ polarized branch is at \textit{higher} energy,
at 4.5T both $\sigma^+$ and $\sigma^-$ emission coincide in energy and
at $B_{z'}>4.5T$ the $\sigma^+$ is finally at \textit{lower} energy. 
So the Zeeman splitting versus $B_{z'}$ is first tending towards negative
values, before decreasing in amplitude to pass through zero
at $B_{z'}^0\simeq 4.5T$ to finally become positive. For dots showing
this reversal in sign for the Zeeman splitting, the exact value of
$B_{z'}^0$ varies from dot to dot.  The Zeeman
  splitting extracted from the spectra following the procedure of
  Ref.~\cite{belhadj09} is plotted in Fig.~2d and clearly demonstrates the change in sign.
 
The evolution of the bright X$^0$ splitting varies dramatically from dot to dot:
For QD II which has at $B_{z'}=0$ very similar emission characteristics to QD I (transition energy, exciton states, values of $g$ factors and exchange energies)
 we record a splitting that is always positive and does not
change sign, see Fig. 2h. The absolute value of the Zeeman splitting
at $B_{z'}=9$ Tesla is a factor of three higher in QD II than in QD I.  
Both dots QD I and QD II show prominent dark state emission
and for both dots
the dark state Zeeman splitting is a monotonous function of $B_{z'}$,
as shown in Fig. 1i for QD I.

\begin{figure}
\includegraphics[width=0.48\textwidth]{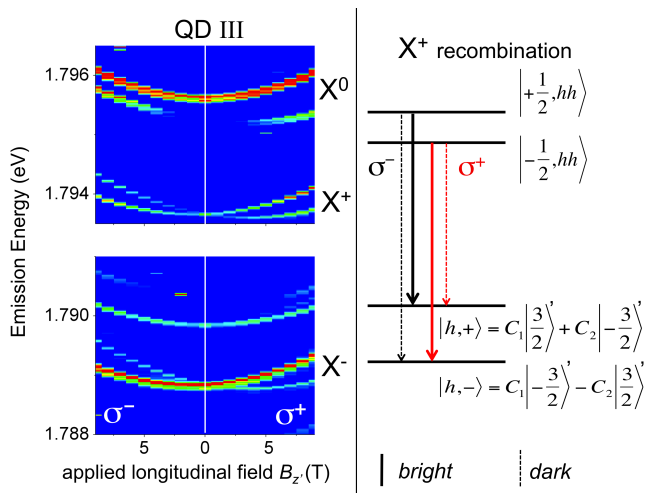}
\caption{\label{fig:fig3} (Color online) Left: Bright and dark states
  for \textbf{QD III} vs $B_{z'}$. Right: Scheme of recombination of X$^+$ with
  mixed hole states, see Eq. \ref{2x2}. \  
}
\end{figure}

At the origin of all these surprising effects lies the magnetic field
induced mixing between the heavy hole states with the angular momentum 
projection $\pm 3/2$ onto the growth axis $z'$.  Let us introduce
the coordinate system $x' \parallel
[11\bar{2}]$, $y' \parallel [\bar{1}10]$ and $z' \parallel [111]$
relevant for the structure under study and
the heavy-hole basis functions $| 3/2 \rangle', |-3/2\rangle'$ which
transform according to the representation $\Gamma_5 + \Gamma_6$, where $\Gamma_{5,6}$ are irreducible
representations of the group C$_{3v}$. It is crucial to note that
the symmetry properties of the field $B_{z'}$ are described by the
representation $\Gamma_2$ and 
the direct product $(\Gamma_5 + \Gamma_6) \times( \Gamma^*_5 +
\Gamma^*_6) = 2\Gamma_1 + 2\Gamma_2$ contains not one, but two
representations $\Gamma_2$. As a result the heavy-hole Zeeman
splitting in the basis $| 3/2 \rangle', |-3/2\rangle'$ is described by
the 2$\times$2 matrix with two linearly independent coefficients 
\small
\begin{equation} \label{2x2}
{\cal H}_{\bm B} = \frac12\ \mu_B B_{z'} \left[ \begin{array}{cc}
g_{h1} & g_{h2}\\  g_{h2} & - g_{h1}\end{array} \right] \:. 
\end{equation}
\normalsize
Here $\mu_B$ is the Bohr magneton, $g_{h1}$ and $g_{h2}$ are the
effective hole $g$-factors. We emphasize
that the above arguments hold for heavy holes in a system of any
dimensionality $n$D $(n=0...3)$ provided its symmetry 
is trigonal, including an exciton formed in bulk Germanium by an electron in
the $L$-valley and a $\Gamma_8^+$ hole.~\cite{averkiev81}  In contrast,
in conventional  [001] grown structures, the longitudinal-field  
induced mixing of heavy holes is symmetry-forbidden, $g_{h2} \equiv
0$~\cite{transverse}. Microscopically, 
a non-zero value of the off-diagonal coefficient $g_{h2}$ in the
[111] grown systems can already be obtained 
within the framework of the bulk hole Zeeman Hamiltonian which in the cubic
axes $x$, $y$, $z$ reads~\cite{Ivchenko1995}
\small
\begin{equation} \label{H_Zeeman}
{\cal H}_{\bm B}
= - 2 \mu_B \left[ \kappa {\bm
J}\cdot{\bm B} + q (J_x^3 B_x + J_y^3 B_y + J_z^3 B_z) \right],
\end{equation}
\normalsize
where  $\kappa$ and $q$ are dimensionless coefficients,  $J_x, J_y$ and $
J_z$ are the angular momentum matrices in the $\Gamma_8$
basis. Transition in Eq.~(\ref{H_Zeeman}) to the coordinate system
$x',y',z'$ gives the relations between the pairs of coefficients in
Eqs.~(\ref{2x2}) and (\ref{H_Zeeman}): $g_{h1}=-[6\kappa + (23/2)q]$,
$g_{h2} = 
2\sqrt{2}q$. In bulk semiconductors the coefficient $q$ is too
small~\cite{Marie1999} to be responsible for high values of
$g_{h2}$. However, in low-dimensional
systems the hole $g$ factors are very sensitive to the strength and
shape of the confining potential. Particularly, an important
contribution to $g_{h2}$ could 
be given by valence-band spin-orbit terms cubic in wave vector
$k$. The relevant contribution $\propto 
J_x^3 k_x (k_y^2-k_z^2)+ \ldots$  to the hole
Hamiltonian~\cite{Rashba1988175} can be recast as 
$ {\cal H}_{v3} = {\cal A}^{(3)} (J_{x'}^3 - 3 \{ J_{x'}J_{y'}^2\})\
{\rm Im}(k_{x'} - {\rm i} k_{y'})^3\ $
where the curly brackets mean the anticommutator \cite{Ivchenko1995}. Replacing ${\bm k}$
by ${\bm k} - (e {\bm A}/c \hbar)$ with ${\bm A}$ being the vector
potential of the magnetic field we obtain 
\small
\begin{equation} \label{ghcubic}
g_{h2} = -18 \frac{m_0 {\cal A}^{(3)}}{\hbar^2} \left\langle \left(
  \frac{\partial}{\partial x'} - {\rm i} \frac{\partial}{\partial y'}
\right)^2 (x'-\mathrm i y') \right\rangle\:,
\end{equation}
\normalsize
where $m_0$ is the free electron mass and the averaging is carried out
over the confined-hole envelope function.

In a longitudinal magnetic field, the hole eigen energies are
$E_{\pm} = \pm g_h \mu_B B_{z'}/2$ with $ g_h = \sqrt{g_{h1}^2 + 
g_{h2}^2}$ and the hole eigenstates $|h,\pm \rangle$ are admixtures
of $|3/2\rangle'$ and $|-3/2\rangle'$, as indicated in Fig.~3, with
the coefficients  
$C_{1,2}$ determined solely by the ratio $g_{h2}/g_{h1}$.  
For non-zero $g_{h2}$, all the four radiative transitions are allowed,
each transition being  
circularly polarized, either $\sigma^+$ or $\sigma^-$~
\cite{footlight}. For illustration, the four channels of radiative  
recombination of a positively charged trion are shown on the
right-hand side of Fig.~3, together 
with the corresponding sign of circular polarization. The transition
energies are determined by combinations of the
electron and hole effective $g$-factors which allows to find a pair of
parameters, $g_e$ and $g_h=\sqrt{ g_{h1}^2 + g_{h2}^2}$. The
intensities of circularly-polarized lines are proportional to
$|C_1|^2$ 
and $|C_2|^2$ and independent of the magnetic field, in full agreement with our
experiments. From the ratio of intensities of identically-polarized lines
we can find the ratio $g_{h1}/g_h$ and, therefore, determine values of
$g_e$, $g_{h1}$ and modulus of $g_{h2}$.

The values of the $g$ factors vary from dot to dot and
even for different complexes X$^0$, X$^+$, X$^-$ in the same dot
revealing the importance of confinement 
and Coulomb interaction for the $g$-factor renormalization. Values for five
typical dots are listed in Table~\ref{tab:table1} \cite{footge}. The
experimental observation of dark 
states for all dots investigated leads logically to
$g_{h2}\ne 0$ for all dots.

\begin{table}
\caption{\label{tab:table1} $g$-factors (typ. error $\leq 10 \%$) for charged
  and neutral excitons obtained from fitting the data. For the X$^0$ the
  $g_e$ and $g_h$ values obtained for X$^+$ for the same dot are taken and only
 $|g_{h2}|$ is varied to fit the bright and dark $X^0$ splitting
  \textit{simultaneously}. } 
\begin{ruledtabular}
\begin{tabular}{cccccc}  
  & QD I & QD II & QD III & QD IV & QD V  \\
\hline
X$^-$ : $g_{e}$ & 0.49 &	0.46 &	0.47 & 0.48 &	0.50 \\
$g_{h}$ & 0.83 &	0.71	& 0.81 & 0.79 & 0.74 \\
$|g_{h2}|$ & 0.53 &	0.60 &	0.53 &	0.57 &	0.57 \\
\hline
X$^+$ : $g_{e}$  & 0.47 &	0.44	& 0.44 &	0.47	& 0.50 \\
$g_{h}$ & 0.71 & 0.72 & 0.72 & 0.72 & 0.73\\
$|g_{h2}|$ & 0.62	& 0.72 & 0.68 & 0.70 & 0.72 \\
\hline
X$^0$ : $|g_{h2}|$ & 0.50 & 0.68 & 0.56 & 0.59 & 0.65 \\
\multicolumn{6}{l}{$g_{e}$ and $g_{h}$: same values as for X$^+$} \\
\end{tabular}
\end{ruledtabular}
\end{table}

Dark transition related to X$^+$ and X$^-$ complexes are always
  present in the spectra for all non-zero values of the field. By
  contrast, emission intensities of dark X$^0$ states increase gradually with $B_{z'}$. This is a result of the effect
  of electron-hole exchange interaction. Taking into account isotropic
  short-range and long-range exchange interaction and assuming that
  the confining potential possesses 3-fold rotation axis we obtain for
  the X$^0$ sublevel energies 
\small
\begin{eqnarray} \label{esm}
E_{s,m} = s g_e \mu_B B_{z'} + \frac12 (\delta_0 + m \delta_m)\:, \hspace{2.5 cm}\mbox{}\\
\delta_{m} = \sqrt{\delta_0^2 + (g_h\mu_B B_{z'})^2 - 4 s g_{h1} \mu_B B_{z'} \delta_0}\:.\nonumber
\end{eqnarray}   
\normalsize 
Hereafter we assume the exchange splitting between bright and dark
states, $\delta_0>0$, $s=\pm 1/2$ denotes electron spin, $m=\pm 1$
denote heavy hole states $|h,\pm\rangle$, respectively. At zero magnetic field the higher sublevels
$(\pm 1/2,+)$ are bright and the lower sublevels $(\pm 1/2,-)$ are
forbidden. The optical activity of
the dark states is induced by the magnetic field in our
experiments \cite{aniso}. Exciton
energy is referred to the zero-field dark state with $m=-1$.
It follows from Eq.~\eqref{esm} that the splitting of bright
X$^0$ states, $E_{+1/2,+}(B_{z'}) - E_{-1/2,+}(B_{z'})$,  can be a
non-monotonous and sign-changing function of $B_{z'}$. This is
confirmed by our measurements shown in Fig. 2d, where the
calculation (solid line) follows closely the experiment (dots). The most surprising feature, the
vanishing X$^0$ splitting at $B_{z'} = B_{z'}^0$ is well reproduced by the model. This
result is another striking 
difference when comparing with the work on [001] grown dots, where the
observed splitting increases monotonously as a function of the applied
longitudinal field \cite{Leger2007,Besombes2000,Bayer1999}.
The   fit of the data in Fig.~2d is very sensitive to the exact value of
$|g_{h2}|$ which explains the strong variations of the X$^0$ bright
splittings as a function of $B_{z'}$ from dot to dot. 
To go from the strongly non-monotonous behavior for QD I 
to the more monotonous graph for QD II  in Fig.~2h, 
a change of $|g_{h2}|$ of only about $20\%$ is sufficient, all other parameters remaining constant.
Here the development of a microscopic theory for
$g_{h1}$ and $g_{h2}$ for realistic quantum dot samples will deepen our understanding. 
  
Taking into account (i) the energy dependence of the bright X$^0$ on
$B_{z'}$, (ii) the polarization of the X$^0$ eigenstates and our
spectral resolution we calculate the emission spectra in the
$\sigma^+/\sigma^-$ basis using the fitted $g$-factor
values. Our theory shown in Fig.~2e-g reproduces the measurements very accurately in terms of
sign and value of the splitting and emission polarization. 
Interestingly, the X$^0$ eigenstates 'exchange' polarization at the
field value $B_{z'}^0$. For $B_{z'}<B_{z'}^0$ the calculations and
measurements show that the higher lying state is $\sigma^+$ polarized
and the lower $\sigma^-$; at $B_{z'}>B_{z'}^0$ it is the opposite. 
Inclusion of small but non-zero anisotropic splitting of
  bright doublet, $\delta_1$, results in the non-vanishing splitting
  of the eigenstates for all values of $B_{z'}$, as shown in Fig.~2d
  by the dash-dotted curve. However, at $B_{z'} \approx 0$ and $B_{z'}^0$
  $\sigma^+$ and $\sigma^-$ polarized lines exchange their places. Our
  measurement scheme allows us to extract the Zeeman splitting
  only~\cite{belhadj09}. The influence of $\delta_1$ and the
  determination of exact polarization eigenstates of the system sets
  the challenge for future experiments, aiming to eventually tune
the X$^0$ splitting to zero to  erase the 'which path'
information, a necessary condition for the generation of entangled
photon pairs from the biexction-exciton cascade
\cite{Stevenson2006,Plumhof2011}. 
Also additional energy shifts due to nuclear spin effects
will be explored in this context. 
  Traces of heavy-hole mixing should also be investigated in [111] grown
GaAs/AlGaAs quantum wells \cite{Vina1992}, currently at the centre of
interest due to the predicted ultra-long electron spin relaxation times \cite{Cartoixa2005}.

We thank ANR QUAMOS, ANR SPINMAN, ITN SPINOPTRONICS, RFBR, LIA CNRS ILNACS
  and Dynasty Foundation for support.


\begin{thebibliography}{31}
\bibitem[{foo({\natexlab{a}})}]{footlight}
\bibinfo{note}{If in addition to the discussed effects heavy-hole to light-hole
  coupling were important, four lines would be observed for both $\sigma^+$ and
  $\sigma^-$ polarized emission.}
\end{thebibliography}
\end{document}